\documentclass[12pt,preprint]{aastex}

\newcommand{\kepler}{\emph{Kepler}}
\newcommand{\tess}{\emph{TESS}}
\newcommand{\minprecision}{300 ppm}
\newcommand{\mediumprecision}{600 ppm}
\newcommand{\numbins}{50}
\newcommand{\binsize}{10-min}
\newcommand{\boxcarwidth}{0.5 days}

\slugcomment{\today}

\shorttitle{Looking for Very Short-Period Planets with Re-Purposed \kepler}
\shortauthors{Jackson \& Stark}

\begin{document}
\title{Looking for Very Short-Period Planets with a Re-Purposed \kepler~Mission}
\author{Brian Jackson}
\affil{Carnegie Institution for Science, 5241 Broad Branch Road, NW, Washington, DC 20015, USA}
\email{bjackson@dtm.ciw.edu}


\begin{abstract}
\indent A re-purposed \kepler~mission could continue the search for $\sim$Earth-sized planets in very short-period ($\le$ 1 day) orbits. Recent surveys of the \kepler~data already available have revealed at least a dozen such planetary candidates, and a more complete and focused survey is likely to reveal more. Given the planets' short orbital periods, building the requisite signal-to-noise to detect the candidates by stacking multiple transits requires a much shorter observational baseline than for longer-period planets, and the transits are likely more robust against the much larger instrumental variations anticipated for the modified \kepler~pointing capabilities. Searching for these unusual planets will also leverage the \kepler~mission's already considerable expertise in planetary transit detection and analysis. These candidates may represent an entirely new class of planet. They may also provide unprecedented insights into planet formation and evolution and sensitive probes for planet-star interactions and the stellar wind. Whatever their origins and natures, such planets would be particularly amenable to discovery by the planned \tess~mission, and a preliminary survey by \kepler~could pave the way for such \tess~discoveries.
\end{abstract}

\section{Description of the Proposed Science Project}
\label{sec:proposed_project}

\begin{figure}
\centering\includegraphics[width=\textwidth]{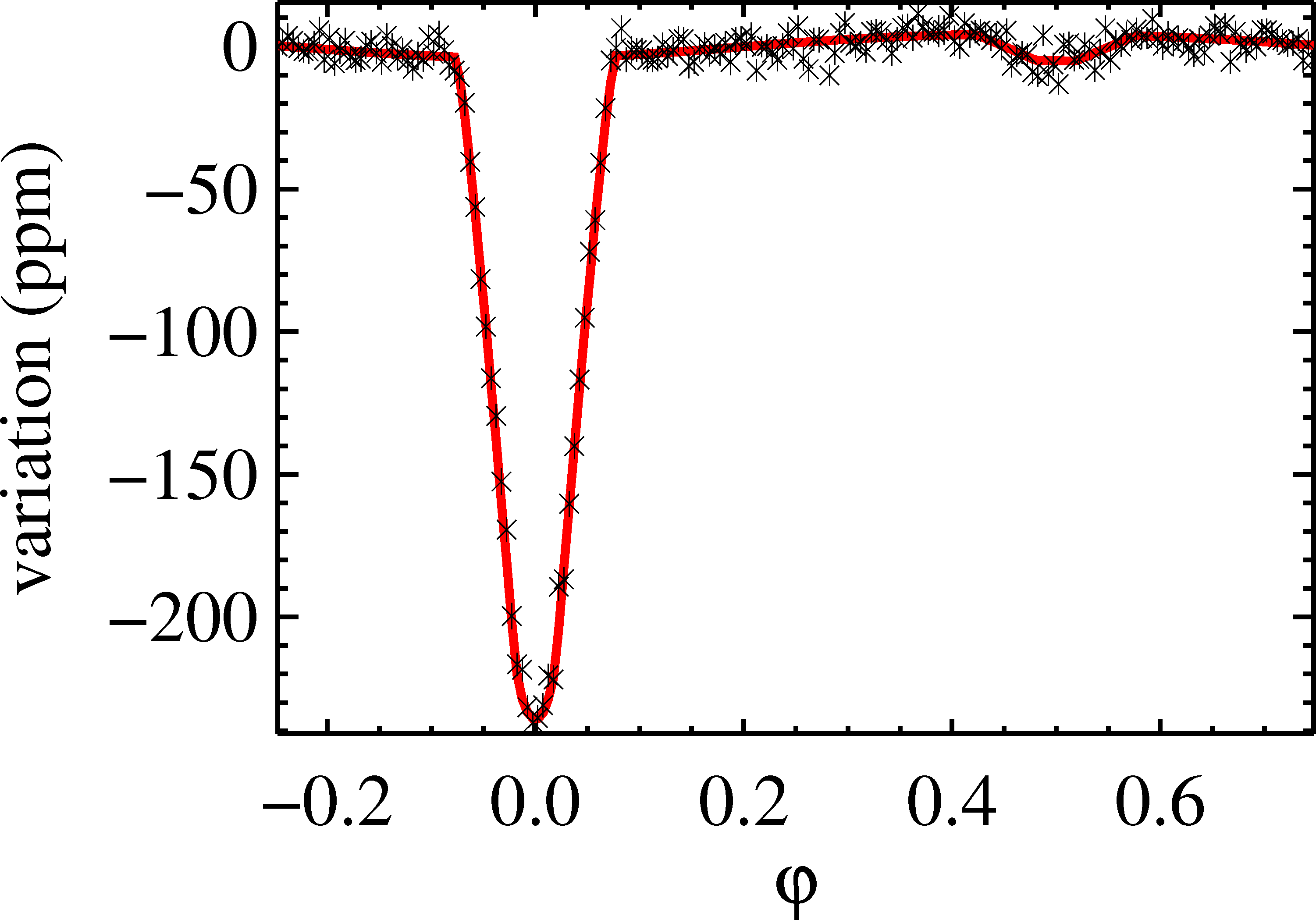}
\caption{Photometric model fit (red line) and data (x's) for KIC 8435766, with a period of 0.3550080 days. $\phi$ is orbital phase ($=$ 0 at mid-transit). Given the typical ingress/egress times for very short-period candidates ($\sim$ 1-min long) and \kepler's 30-min long-cadence, the transits can be shallower and more v-shaped than usual planetary transits.}
\label{fig:KIC8435766_fit_EVs_MCMC}
\end{figure}

\indent Almost every new discovery in extrasolar planetary (exoplanetary) astronomy challenges a previously-held truth about the origins and natures of planets. In particular, the discoveries of hundreds of exoplanets orbiting within 0.1 AU of their host stars with orbital periods of a few days (close-in planets) have defied conventional planet formation and evolution models based on our solar system. Recent studies have shown that there are smaller planetary candidates in even closer orbits than hot Jupiters. For example, \citet{2013ApJ...774...54S} announced the discovery of Kepler-78 b (orbiting KIC 8435766), which has an orbital period of about 0.355 days or 8.5 hours. In one \kepler~quarter, the candidate transited its star more than \emph{250 times}. The planet's transit depth is about 200 ppm, so for a photometric precision of \minprecision, a $>$ 6-$\sigma$ detection of this planet's transit requires less than one \kepler~quarter of observations. \citet{2013arXiv1308.1379J} independently discovered this candidate, and Figure \ref{fig:KIC8435766_fit_EVs_MCMC} shows the binned transit data for Kepler-78 b from that study.

\indent These candidate planets await confirmation via follow-up observations (with a precision $\sim$ 100 m/s, \citealt{2013ApJ...774...54S} reported no RV signal for Kepler-78 b, bolstering the planetary interpretation). However, given their proximity to their host stars, such planetary candidates are ideally suited for RV follow-up. For example, a 10-$M_\mathrm{Earth}$ planet with an orbital period of 5 hours induces an RV signal of 10 m/s in a solar-mass star. Sufficient  precision to detect this signal is currently achievable from ground-based instruments, so continued discovery of such candidates should be a high priority. 

\indent In the following white paper, I discuss the possibility of a continued search for very short-period, Earth-sized planets using a re-purposed \kepler~spacecraft. In Section \ref{sec:questions}, I answer the specific technical questions posed by the white paper solicitation, and in Section \ref{sec:model}, I present a  simplified model for detection of hypothetical transit signals in synthetic data. In Section \ref{sec:impact}, I briefly describe the anticipated scientific impact of this project.

\section{\kepler~White Paper Solicitation Questions}
\label{sec:questions}

\indent I leave unanswered a whole host of astrophysical and technical questions about the proposed survey, but it is very similar to the transit search conducted during the nominal \kepler~mission and could even be conducted contemporaneously with other studies, assuming suitable target stars can be identified in the field of view. 

\begin{itemize}
\item {\bf How will the focal plane will be used?} -- Target apertures -- The expected capabilities during a re-purposed \kepler~mission can complicate target aperture observations, and rather than using small clusters of pixels for each target (as in the nominal mission), perhaps whole regions of \kepler's CCD can be selected. For example, if target image's trajectory on the CCD during an observation can be predicted, then perhaps data from all pixels along the trajectory can be returned. The position of the target photocenter and the pixel response function can be solved for afterward. Obviously, the capabilities of the detector and the available bandwidth may limit such a strategy.

\item {\bf Planned integration time(s)} -- Probably the shorter, the better -- The expected pointing drift will allow more the target image to sample more pixels, which will likely increase the noise during one integration from non-uniform pixel response. One strategy to mitigate this effect might be to de-focus the telescope somewhat (although de-focusing may complicate other, contemporaneous observations). Also, because the in/egress times for very short-period transiting planets are so short ($\sim$ minutes), long integration times make the transits more shallow, which can complicate the transit light curve analysis and make distinguishing between planetary and binary star occulations on the basis of light curve shape more difficult. However, the integration times must be long enough that sufficient photons are recovered for transit detection in the first place. The lower limit for integration time will depend on the target stars' brightness.

\item {\bf Expected data storage need} -- TBD.

\item {\bf Data reduction or analysis plans} -- Planning the details of data reduction requires knowing the instrument's performance and the sources of noise better. Presumably, though, in searching for more (but shorter than previously) planetary transits, the data reduction strategy would closely resemble that used during the nominal mission. Data analysis would proceed as described in Section \ref{sec:model} and \citet{2013arXiv1308.1379J}.

\item {\bf What class of science target is involved} -- Point sources.

\item {\bf Target durations} -- Probably observations over several days to weeks, and significant drift could still allow recovery of the transits, depending on the noise (Section \ref{sec:model}).

\item {\bf How long the science program should be run} -- TBD.

\item {\bf What scientific impact will occur during and after the proposed project} -- See Section \ref{sec:impact}.

\end{itemize}

\section{Model Data}
\label{sec:model}


\indent To demonstrate the recovery of the transit of a very short-period planet, I generated synthetic transit data, incorporating both white and red noise. Using the transit light curve formalism described in \citet{2002ApJ...580L.171M} and \citet{2013PASP..125...83E}, I first generated two sets of synthetic transit data, considering the central transit of a 1.5 Earth radius $R_\mathrm{Earth}$ planet across the Sun. I assumed orbital periods $P = 0.355008$ and $2 \times 0.355008 = 0.710016$ days. These transits have durations of 1.5 and 3 hours, respectively.

\begin{figure}
\centering\includegraphics[width=\textwidth]{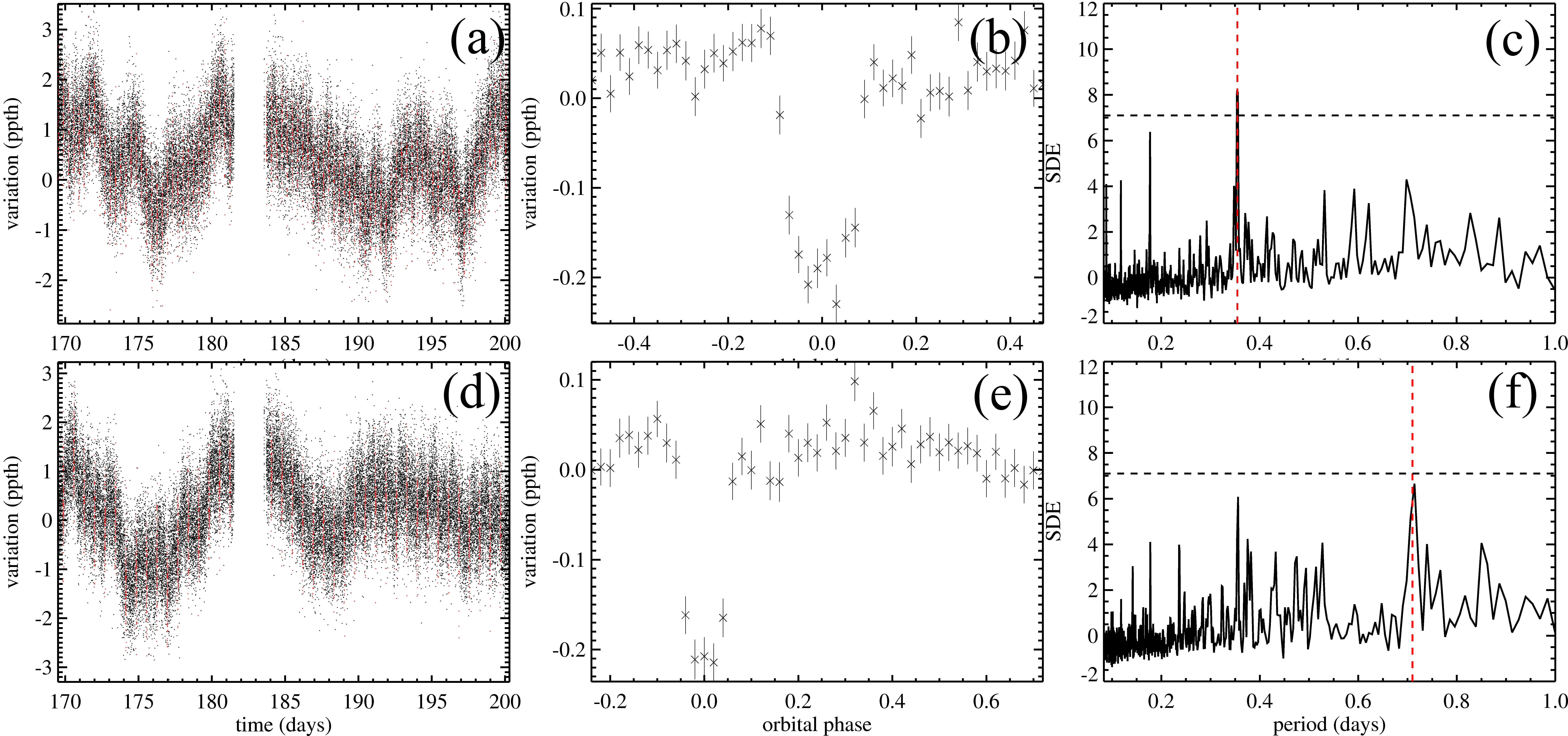}

\caption{Example transit recovery. For $P = 0.355008$ days, panels (a)-(c) show unfolded data (in parts-per-thousand, ppth), data folded on the period and binned into \binsize~bins, and an EEBLS spectrum (with the correct period highlighted by the red line), respectively. Panels (d)-(f) show the same, except for $P = 2 \times 0.355008 = 0.710016$ days. In (a) and (d), red dots indicate the points in transit. In the EEBLS spectra, the horizontal dashed line shows a 7.1-$\sigma$ detection threshold.}
\label{fig:recover_example_transits_red}
\end{figure}

\indent I took the white and red noise both to have standard deviations of \mediumprecision. To generate the red noise, I first generated white noise and smoothed it (using IDL's SMOOTH function) with a smoothing length corresponding to 4 days since the solicitation indicated there will be significant drift in the telescope pointing over a timescale $\sim$ 4 days -- presumably this drift will lead to systematic variations in the measured fluxes on this timescale. I generated 30 days (1/3 of a \kepler~quarter) of data using an observational cadence of 1-min. Over this baseline, our model planet transits its star about 84 and 42 times with the shorter and longer periods, respectively. Panels (a) and (d) from Figure \ref{fig:recover_example_transits_red} show these synthetic data.

\indent I then attempted to recover the transits. To remove the red noise, I applied a mean boxcar filter with a window width of \boxcarwidth, which introduces negligible distortion of the transit signals. Panels (b) and (e) show the resulting transit curves after folding the detrended data on the known orbital periods and binning into \numbins~bins -- the uncertainty on each datum is about 20 ppm, about 10 times smaller than the transit depth. I applied the EEBLS algorithm \citep{2002A&A...391..369K} and calculated the Signal Residue $SR$, which is maximized when the algorithm finds the right transit period. For each trial period, EEBLS reports the maximum $SR$ from among all possible transit phases and all given trial durations. From the $SR$-values, I calculated the $SDE$-values, defined as $SDE = (SR - \langle SR \rangle)/\sigma(SR)$, where $\langle SR \rangle$ is the mean $SR$ and $\sigma(SR)$ its standard deviation. When the noise is white, $SDE$ is a Gaussian random variable, so larger values are increasingly unlikely in the absence of a transit signal. Panels (c) and (f) from Figure \ref{fig:recover_example_transits_red} show discernible peaks at the assumed periods (and period aliases), although the peak is less pronounced for the longer period. In fact, the longer-period transit is only marginally recovered.

\begin{figure}
\centering\includegraphics[width=\textwidth]{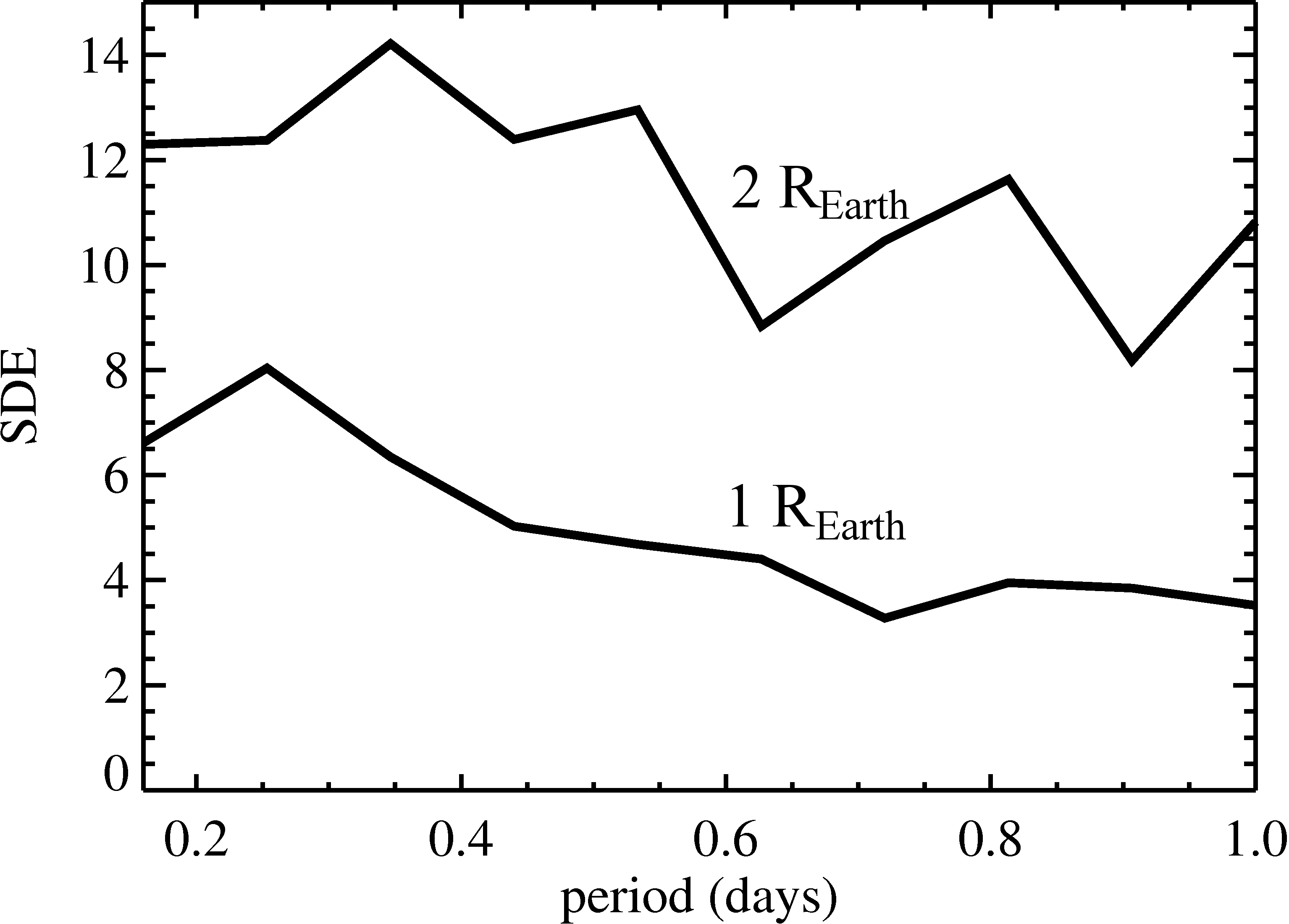}
\caption{SDE for transits in the presence of red noise with a correlation time of 4 days for a range of orbital periods and radii (as indicated by the line labels).}
\label{fig:SDE_vs_period}
\end{figure}

\indent To investigate the dependence of transit recovery on orbital period $P$, I generated a grid of synthetic transit data with the same noise characteristics as above for a range of planetary radii and orbital periods and detrended them in the same way. Figure \ref{fig:SDE_vs_period} shows $SDE$ as a function of orbital period and transit radius. (Jaggedness in the figure arises from statistical noise that future, planned analysis will tame.) For the range of periods shown, $SDE$ is mostly sensitive to transit radius but still somewhat sensitive to period, and the sensitivity increases for even longer periods than those shown, particularly as the orbital period approaches the correlation timescale for the red noise. 

\begin{figure}
\centering\includegraphics[width=\textwidth]{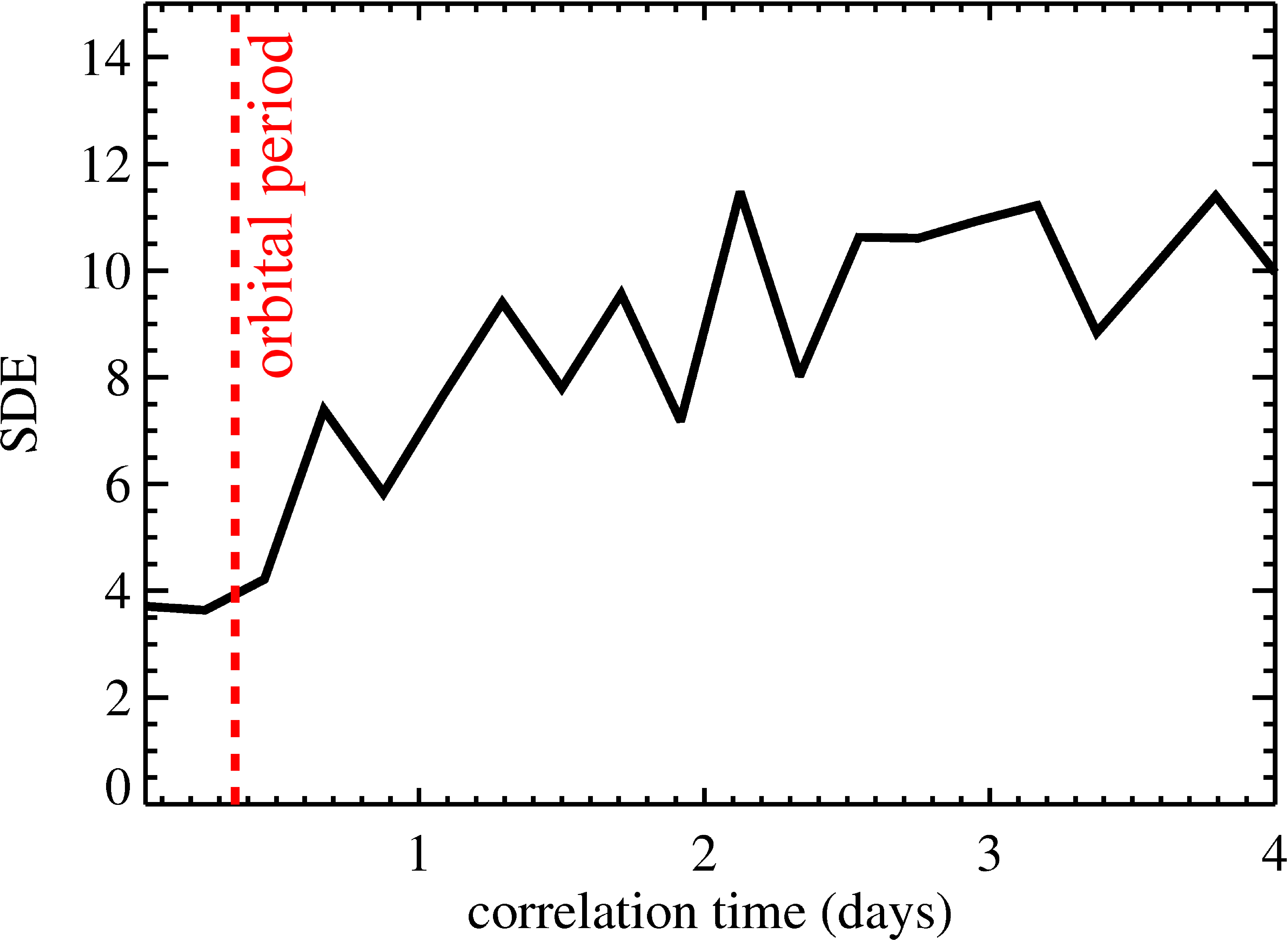}
\caption{SDE for detecting transits of a 1.5 $R_\mathrm{Earth}$ planet transiting the Sun in a 0.355008-day orbit, in the presence of red noise with a range of correlation times.}
\label{fig:SDE_vs_corr_time}
\end{figure}

\indent I also investigated transit recovery for a range of correlation timescales. For this calculation, I assumed all the same transit, data, and data conditioning parameters as above, with $P = 0.355008$ days and a transit radius of 1.5 $R_\mathrm{Earth}$, but considered red noise correlation timescales from 1 hour (0.042 days) up to 4 days. Figure \ref{fig:SDE_vs_corr_time} shows the $SDE$-values' dependence on the correlation timescale. 

\indent As the correlation time surpasses the orbital period, the transit is more robustly recovered. Potentially, detrending the data with a smaller boxcar window could mitigate the red noise with the shortest correlation timescales, but detrending with a window of a width comparable to the transit duration can distort the transit beyond detectability. This calculation illustrates one major advantage of looking for transits of very short-period planets: In addition to allowing significant signal-to-noise to build over a relatively short observational baseline, the very short-period transits can be effectively recovered, even with fairly short timescale (many hours) systematic variations that will likely arise from pointing drift.

\section{Scientific Impact and Relevance to NASA Astrophysics Program}
\label{sec:impact}

\indent This proposed project will have many of the same scientific impacts as the nominal \kepler, except for discovering habitable extrasolar planets. For example, Kepler-78 b has an equilibrium day side blackbody temperature $\sim$ 2,300 K, probably hot enough to preclude (Earth-like) life. However, the proposed survey clearly addresses questions that emanate from NASA SMD's Astrophysics goals, as discussed on p.~60 of the 2010 SMD science plan: ``How did the universe originate and evolve to produce the galaxies, stars, and planets we see today?'' and ``What are the characteristics of planetary systems orbiting other stars?'' Although as yet unconfirmed, searching for this novel class of planet candidate may elucidate several important astrophysical processes and pave the way for \tess:

\begin{itemize}
\item In addition to many other planet types, the upcoming \tess~mission will be ideally suited for finding more planets with very short-periods. If \kepler~can get a head start discovering these kinds of planets, it can help guide \tess, perhaps helping designate interesting targets. Also, a long observational baseline extending between missions may allow detection of the slow orbital decay resulting from tidal interactions between the planets and their host stars \citep{2013arXiv1308.1379J}.

\item These very short-period planets may probe planet-star interactions and stellar magnetospheres. Several studies give tantalizing evidence for planet-stellar magnetosphere interactions \citep{2010EGUGA..1213591S}. With surface temperatures $>$ 2000 K, these short-period planets may shed rock vapor atmospheres \citep{2013MNRAS.433.2294P}, and the interaction between the atmospheres and stellar wind may be observable, as for Mercury \citep{2008merc.book..251K}. Although such detection would probably require a significant investment of observational facilities, detection of such an atmosphere could provide direct constraints on the compositions of very short-period planets. 

\item The origins of very short-period planets are still unclear (if they \emph{are} planets), but \citet{2013arXiv1308.1379J} suggested they might be the fossil cores of disrupted hot Jupiters. Although this hypothesis is far from confirmed, detection of additional very short-period planets would help shed light on their origins, and if they can be identified as fossil cores, the planets would provide unprecedented insights into gas giant formation. For example, for gas giants that originate through core accretion, the threshold core mass that initiates atmospheric accretion is unconstrained by observation and probably depends on protoplanetary conditions, such as metallicity \citep{2006ApJ...648..666R}. 

\end{itemize}


\vspace{-18pt}

\acknowledgments
\indent BJ acknowledges helpful input from Alan Boss, Joleen Carlberg, Drake Deming, and Christopher Stark.

\vspace{-30pt}

\bibliography{Jackson_Stark_kep-wp_refs}

\end{document}